# Examining the drivers of business cycle divergence between Euro Area and Romania

**Ionuț JIANU**
Bucharest University of Economic Studies, Romania
ionutjianu91@yahoo.com

**Abstract.** *This research aims to provide an explanatory analyses of the business cycles divergence between Euro Area and Romania, respectively its drivers, since the synchronisation of output-gaps is one of the most important topic in the context of a potential EMU accession. According to the estimates, output-gaps synchronisation entered on a downward path in the subperiod 2010-2017, compared to 2002-2009. The paper demonstrates there is a negative relationship between business cycles divergence and three factors (economic structure convergence, wage structure convergence and economic openness), but also a positive relationship between it and its autoregressive term, respectively the GDP per capita convergence.*

## 1. Introduction

One of the fundamental reasons for establishing single currencies is related to the political actions aimed at the depreciation of the national currency, which were frequently used by the national governments to enhance the economic growth rate. However, such a strategy was not beneficial, as it encouraged a similar political reaction from other neighboring countries, leading to a general competitive devaluation on medium and long-run. Euro currency was the response of the European Union to this and other challenges. However, the process of adopting euro is quite difficult and, apart from meeting the nominal convergence criteria set out within the Maastricht Treaty before ERM II accession, and maintaining appropriate economic evolutions in line with these for at least two years (EA accession also needs the support of the ECOFIN Council), it also requires achieving an optimum level of real and structural convergence.

After adopting euro, the Member States lose the independence of the national monetary policy which lowers the number of instruments that the country have at disposal for stimulating economic growth. Following the loss of monetary policy independence, the decision to change the interest rate remains at the discretion of the European Central Bank, depending on the evolution of the euro area economy. In this context, if the euro area economy registers a positive output-gap, while one member of it experiences a recessionary output-gap, the ECB's decision will not be in the benefit of the mentioned country, since raising the interest rate will have a negative effect on this economy (the example remains valid even when the roles are reversed). Therefore, the creation of a common destiny is encouraged by increasing business cycles (output-gaps) convergence. This is also in line with the explanation provided by Mundell (1961) regarding the importance of business cycles synchronisation in ensuring an Optimum Currency Area.

The motivation for choosing this thematic area consists in the important role of the output-gaps convergence with the Euro Area, in the preparation of the euro adoption process. Although business cycles convergence is a major topic of interest for researchers, some results are evasive and do not threat, in a comprehensive manner the full implications of the drivers of the business cycles convergence. In addition, most of the studies provide a general view at European Union/Euro Area level and do not focus on individual cases.

The general objective of this paper is to identify the factors that could enhance/reduce the business cycle convergence/divergence between the Romanian economy and Euro Area and its associated effects, this being achieved by reaching the following specific objectives:

- examining the link between the economic structure convergence and business cycles divergence between Romania and Euro Area;
- identifying the impact of economic openness on the business cycles divergence between Romania and Euro Area;
- examining the link between the wage structure convergence and business cycles divergence between Romania and Euro Area;
- identifying the impact of Romanian GDP per capita convergence towards the Euro Area average on the business cycles divergence between the parts reviewed.

## 2. Literature review

Business cycles synchronisation concept was first introduced in the economic literature by Burns and Mitchel (1946) who considered this as a proxy for economic activities volatilities – catched by the dynamic of GDP. Following that, many economists focused their attention on studying business cycles convergence/synchronisation determinants at European Union level. However, the results depends on the methodology and data used. Catching business cycles convergence/divergence or synchronisation may be difficult, since this indicator is based on output-gap estimates which are extremely volatile depending on the period used, but there are several studies proposing different methodologies. Business cycles estimates are generally influenced by the detrending methods applied, Hodrick-Prescott (1980), Baxter-King (1995) and Christiano-Fitzgerald (2003) filters being the most ones used. According to Artis and Zhang (1995), the simplest way to calculate the synchronisation of business cycles consists in the GDP detrending, followed by the computation of the bilateral cross-correlations using Spearman correlation coefficient. There are other options too, such as the Pearson correlation on periods and subperiods (Flood and Rose, 2010), but this paper analyses the divergence between business cycles as the absolute differences between these output-gaps, as this approach is more suitable to the objective of the study – taking also into consideration its limits, since even close business cycles can be in different phases.

Frankel and Rose (1998) analysed the endogeneity of the Optimum Currency Area and argued that there is a causal relationship in terms of trade integration and business cycles convergence. According to the theory of endogeneity, the two authors showed that the differences between countries are smaller the higher is the level of integration. Their research highlights a positive reaction of the symmetrical shocks across the Euro Area to increasing economic integration.

Other authors demonstrated the positive influence of increasing trade intensity on business cycle synchronization for industrialised countries (Fatás, 1997; Clark and van Wincoop, 2001; Imbs, 2004; Calderon et al., 2007). The importance of trade in driving business cycles convergence was also confirmed by other authors (Böwer and Guillemineau, 2006; Garcia-Herrero and Ruiz, 2008; Lee, 2010; Dées and Zorell, 2011). However, Marinaș (2006) stated that increasing the degree of economic openness is not enough for a country as Romania and, in this context, great attention should also be paid to the similarity with the Euro Area in terms of the structure of exports. Fidrmuc (2004) investigated the determinants of business cycle synchronisation between Visegrád Group countries (Czech Republic, Poland, Slovakia and Hungary) and found that the parameter of the bilateral trade intensity index changes its value and statistical significance in case of insertion of other variables (there were only one variable that remained significant in all cases – intra-industrial trade).

The statistical significance of both trade and economic specialisation was also confirmed by Trăistaru (2004). Imbs (2004) identified a positive relationship between economic structure convergence and business cycles synchronisation. According to the results of the paper, countries response to shocks may converge if the economic structures of the states are similar. This evidence was also supported by other authors (Calderon et al., 2007; Beck, 2013, Kalemli-Ozcan et al., 2001, Siedschlag, 2010), while other authors (Baxter and Kouparitsas, 2005; Inklaar, Jong-A-Pin and de Haan, 2008) have come to a different conclusion which highlights the ambiguity of this relationship, but the most opinions support a positive relationship between these variables. On the other hand, cyclical similarity has proven to be a function in response to the evolution of other variables, such as: fiscal similarity, membership in the customs union, the absolute difference in the long-term interest rate.

From another point of view, some researchers (Massmann and Mitchell, 2004; Cancelo, 2012; Gogas, 2013) demonstrated that business cycles synchronisation could be driven by the EMU accession, while others did not support this hypothesis (Camacho et al., 2006; Mink et al., 2007). Other studies (Kose et al., 2003; Cerqueira and Martins, 2009) have shown that financial openness is another significant factor.

## 3. Methodology

This section describes the main econometric tools and methods used to analyse the drivers of business cycles divergence between Romania and Euro Area over the period 2002-2017. In this context, it worth to be mentioned that I have used Eurostat data with quarterly frequency in order to increase the number of observations, implicitly the robustness of the analysis.

The analysis was performed using Eviews 9.0 software, respectively the Least Squares method in time series window. However, having in mind the risk for heteroskedasticity, I opted for a logarithmic transformation for all statistical data used before applying the estimation method on the following equation:

$$\log(OGdiv)_t = \alpha_0 + \beta_0 \log(OGdiv)_{t-1} + \beta_1 \log(ecstructureconv)_t +$$
$$+ \beta_2 \log(openness)_t + \beta_3 \log(wagestructureconv)_t +$$
$$+ \beta_4 log(realgdpcapconv)_t + \varepsilon_t \quad (1)$$

where:

- *OGdiv* represents the business cycle divergence calculated as the absolute difference between the output-gap (the share of the difference between the real and potential GDP expressed in million euros in the potential GDP) registered by Romania and the one of Euro Area. Even if most of the papers use multiple correlations between countries, this approach cannot be used in time-series data when studying the link between business cycles convergence and its determinants as a consequence of the low number of observations. Therefore, I calculated the absolute difference between the cycles as a proxy for the business cycles divergence. First, the estimation of the OG was performed using the seasonally and calendar adjusted data for the real GDP having 2010 as base year. Further, I have applied the Hodrick-Prescott filter for supporting the decomposition of the data into a non-stationary trend and a stationary cyclical component. In order to strengthen the efficiency of the estimation, I set the lambda value to 1600, its role being to counteract the acceleration of the trend relative to the cyclical component of the Gross Domestic Product.

A negative value of the business cycle captures the existence of a recessionary output-gap, while a positive value records an expansionary phase of it. Hodrick-Prescott filter is based on minimising a function that captures the deviations of the Gross Domestic Product from the potential level and the changes of the growth rate of the trend, this being calculated according to the following formula:

$$\mathrm{HP} = \min_{y_t^*}\left\{\sum_{t=1}^{T}(y_t - y_t^*)^2 + \lambda\sum_{t=1}^{T}[(y_t^* - y_{t-1}^*) - (y_{t-1}^* - y_{t-2}^*)]^2\right\} \quad (2)$$

where $(y_t - y_t^*)$ is the business cycle of the Y country and λ represent the lambda coefficient.

However, I have added the autoregressive term, since the variables proved to be stationary at I(0) and I(1) according to the results provided by the Augmented Dickey-Fuller test, its corresponding lag being selected using the Schwarz Information Criterion.

- *ecstructureconv* is the convergence between the economic structure of the Romanian economy and that of the Euro Area, calculated using the methodological guidance provided by Krugman (1993), as follows:

$$ecstructureconv = 1 - \left(\sum_{z=1}^{Z} abs\,(GVA_{RO} - GVA_{EA})\right) \quad (3)$$

where Z represents all number of economic sectors analysed = 11 (z being one of these) and GVA is the share of the Gross Value Added generated by the sector z in the total.

- *openness* is the degree of the economic openness calculated using the following formula:

$$openness = \left(\frac{imports + exports}{GDP}\right) x\ 100 \quad (4)$$

- *wagestructureconv* represents the convergence between the wage structure of the Romanian economy and that of the Euro Area, its calculation being adapted to the Krugman index presented above.

$$wagestructureconv = 1 - \left(\sum_{z=1}^{Z} abs\,(W_{RO} - W_{EA})\right) \quad (5)$$

where W represents the share of the wages provided in the sector z in total wages per economy.

- *realgdpcapconv* is the share of the Romanian real GDP per capita (expressed in euros) in the one registered by the Euro Area. Since the data for this indicator were not available in its seasonally adjusted form, I used the Tramo-Seats tool in order to increase the feasibility of the data by excluding the seasonal influence.

In addition, I have also used Pearson correlation to examine the business cycle synchronisation between Romania and Euro Area by subperiods (pre-crisis period: 2002-2009 and post-crisis period: 2010-2017), as well as throughout the analysed period.

$$Pearson_{(RO,EA)} = \frac{cov(RO,EA)}{\sigma_{RO}\sigma_{EA}} \quad (6)$$

where *cov(RO, EA)* represents the covariance between the cycles of Romania and Euro Area, and $\sigma$ is the standard deviation.

Further, the reliability of the estimations was checked by testing the following hypothesis: (i) statistical validity of the model – Fisher test; (ii) normal distribution of the residuals – Jarque-Bera test; (iii) autocorrelation of the residuals – Breusch-Godfrey test; (iv) homoskedasticity – Breusch-Pagan-Godfrey; (v) stability of the model – CUSUM test; (vi) existence of the multicollinearity – Variance Inflation Factors test. I used this approach, since these are the main conditions identified by Gauss-Markov for confirming the maximum verisimilitude of the estimators.

## 4. Results and interpretations

This section analyses the main results of the computed estimation. First, I have estimated the output-gap (% of potential GDP) at the level of Romania and Euro Area covering the period 2002-2017. The data evidences that Romania have registered a higher volatility in terms of its specific business cycle, this being quite pronounced in the period 2008-2009. As it can be seen in Figure 1, the Romanian economy had followed an overheating trend in 2007-2008, succeeded by a severe recessionary output-gap. However, this evolution had been also followed by the euro area economy, but the magnitude of overheating was higher in Romania as a consequence of the pro-cyclical policies and the lack of building fiscal buffers. In the period 2010-2017, Romania has entered on a downward path in terms of business cycle synchronisation with Euro Area which is also reflected in Table 1. Following the computation of the Pearson correlation between EA business cycle and the one of Romania, I have found a correlation of 59.09% over the full period analysed (2002-2017). Nevertheless, the data shows a higher correlation of the cycles in the period 2002-2009 (67.1%) then the one calculated for the period 2010-2017 (30.1%). One reason is that there was a gap between the starting point of the crisis in Romania and the one of Euro Area since the linkages between Member States using euro currency are stronger through investment and trade channels. In the third and fourth quarter of 2009, the fall of the potential GDP was higher than the one of the real GDP in Romania, which led to a positive output-gap in these period even if the real GDP dropped with 6.2% in 2009Q3 and 4.0% in 2009Q4. This is also close the AMECO reported data for this indicator, which estimated an output-gap of -0.2% of potential GDP in 2009. The crisis effects in Romania have not been addressed through preventive measures and the corrective ones were strictly oriented to several cuts in government spending which have stifled the economy even more, while EA Member States have responded to the crisis momentum in an integrated and coordinated manner. However, when Euro Area economy started to redress, its cyclical position had been affected by the debt crisis. There was a mix between heterogeneous government actions, different crisis starting point and effects which changed the upward path of the business cycle synchronisation between Romania and Euro Area into a downward one.

**Figure 1.** *Business cycles comparison between Romania and Euro Area*

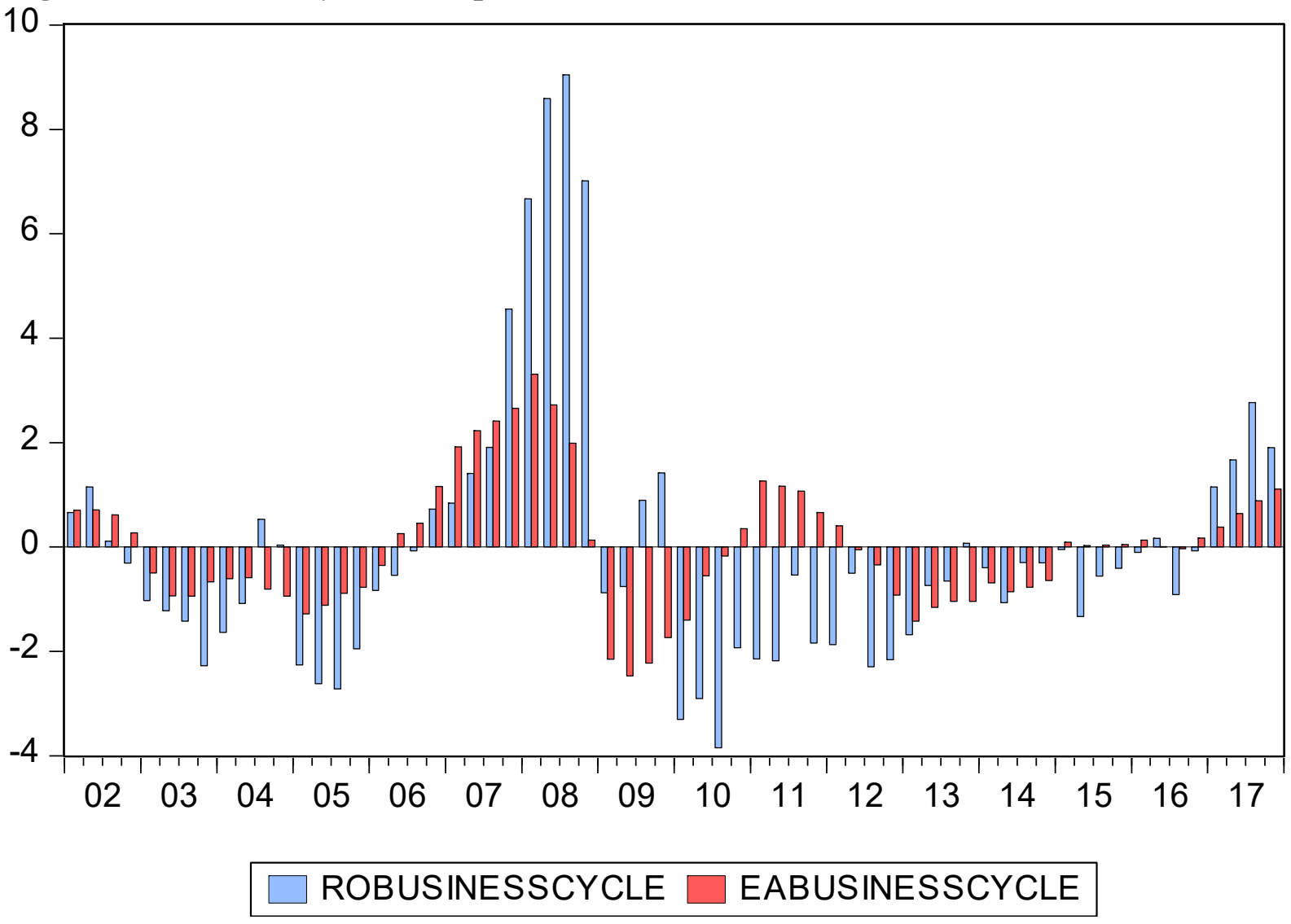

**Source:** Own calculations using Eviews 9.0.

**Table 1**. *Pearson correlation*

| Pearson correlation | Period | | |
|---|---|---|---|
| | 2002Q1-2009Q4 | 2010Q1-2017Q4 | 2002Q1-2017Q4 |
| RO-EA | 67.10% | 30.10% | 59.09% |

**Source:** Own calculations using Eviews 9.0.

Anyway, business cycle synchronisation cannot be used in a time-series model, since it is calculated based on the correlation on a given period, which significantly reduce the number of observations. This argued the need to estimate an indicator that provide an usable value for each quarter of the analysed period. In this context, I have used the absolute difference between the output-gaps in order to catch the estimated values for business cycles divergence.

Figure 2 indicates a stronger inverse relationship between economic structure convergence and business cycles divergence which also support the hypothesis that the first mentioned variable is one the most important driver of the output-gaps convergence. However, this discussion is very tricky since Euro Area economy is more dependent on the service sector and Romania have a higher share of Gross Value Added from the industrial sector in GDP (23.7% of GDP in Romania – 2017, compared to 17.8% of GDP at EA level). Practically, an economy with a high share of the industrial sector could be more resilient in a crisis momentum, while an economy based on the service sector may reflect more economic opportunities and higher profitability. In sum, both examples have their own benefits. Nevertheless, when starting the reform to adopt euro, a country government should start stimulate the economic activity of the competitive sectors, taking also into consideration the economic structure of the Euro Area.

**Figure 2.** *The evolution of business cycles divergence, economic structure convergence and wage structure convergence (RO-EA)*

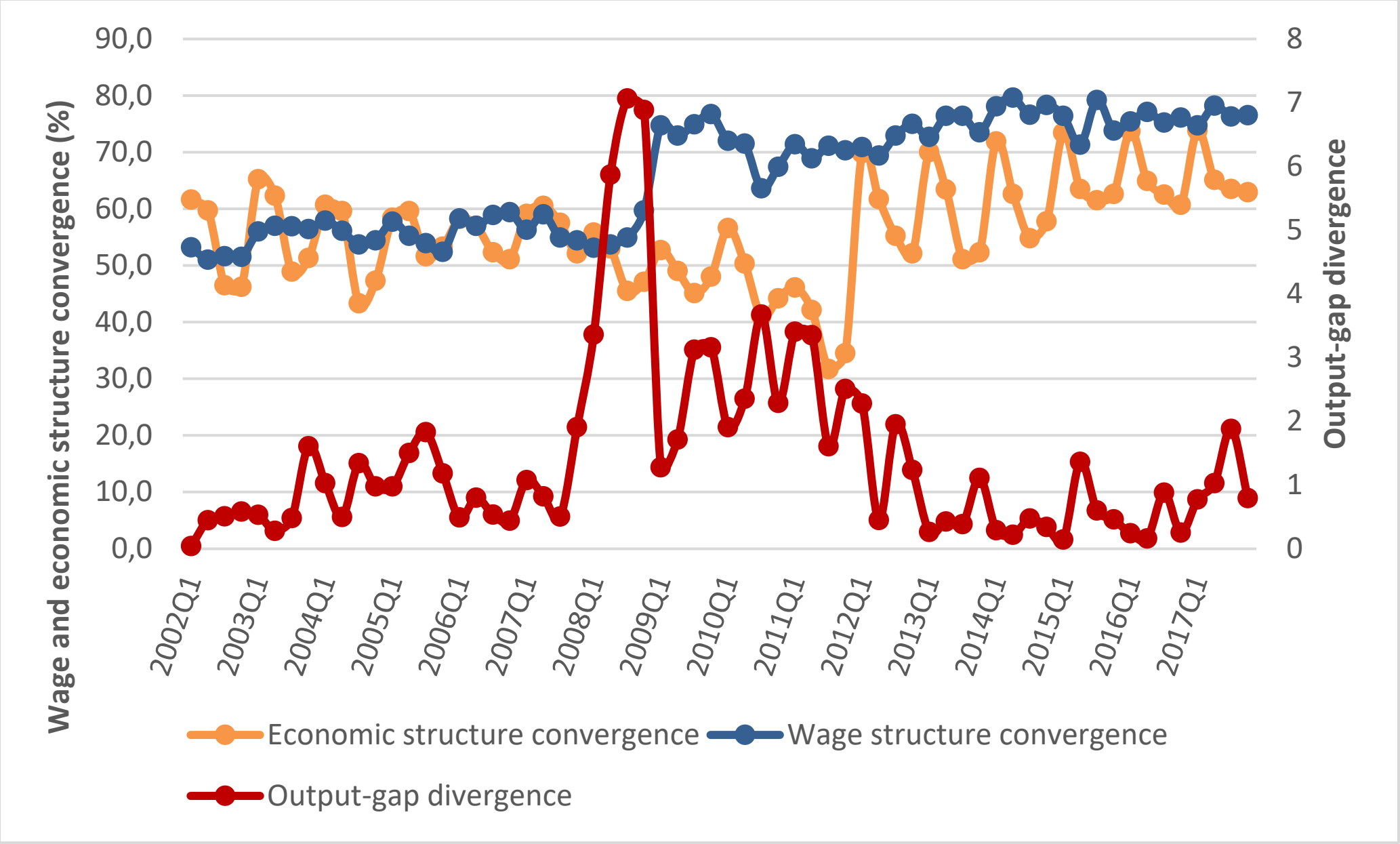

**Source:** Own calculations using Eviews 9.0 and Microsoft Office Excel 2016.

The second studied relationship is that between wage structure convergence and output-gaps divergence which may seen also negative, but lower than the one presented above as a result of the constant evolution of the wage structure convergence which is mainly driven by the wage rigidities. An important evidence is reflected by the results of the statistical correlation between these variables, which indicates a higher negative correlation between business cycles divergence and economic structure convergence (-48.44%) than the one between business cycles divergence and wage structure convergence (-15.65%).

The second phase of the research focuses on estimating the impact of the factors mentioned in methodology on the natural logarithm of output-gaps divergence. In this respect, I found that an increase with 1% in the output-gaps divergence lagged by one quarter (defined in the methodology) led to a rise in the actual business cycles divergence with 0.26%. This can be explained by the facts presented above regarding the different behaviour of the Member States in applying preventive or corrective measures in the starting phases of the crisis and the following ones, when Euro Area have faced with the debt crisis.

Regarding the economic structure convergence between EA and RO, I demonstrated that an increase in its dynamics with 1% generated a fall of output-gaps divergence with 2.18%, which actually represents an increase in business cycles convergence. The main argument supporting this evidence consists in the fact that countries based on kindred economic activities also respond similarly to shocks. Next, I have also identified a negative relationship between the degree of openness and output-gaps divergence, since trade and investment channels creates solid dependent networks, but also exposures between economic players. In fact, according to the estimation, an increase in the degree of openness with 1% have a negative impact on output-gaps divergence of 2.21%.

With a view to the wage structure convergence, I have found that an increase in its dynamics with 1% have led to a fall in the output-gaps divergence of 3.04%. Lastly, I have identified a positive relationship between the real GDP per capita convergence (as it had been defined in the methodology) and output-gaps divergence, since the catching-up process its faster when some EA countries are registering unfavourable developments of potential GDP and Romania reach its peak – which actually places the regions review in different phases of the economic cycles. Therefore, in some cases, this effect may limit the business cycles synchronisation since Romania is a country that need to converge to the EA/EU average in terms of economic development, but prioritisation of the competitive economic activities, respectively the improvement of foreign direct investments level and of the institutions quality could facilitate a balance development favourable to both indicators. Bodislav recommended several actions to promote a sustainable catching-up process, as follows: (i) supporting the increase GDP per capita; (ii) diversifying the structure of the production sectors; (iii) increasing trade and financial openness; (iv) increasing the competitiveness of labour force.

**Figure 3.** *Results of the model*

Dependent Variable: LOG(OGDIV)
Method: Least Squares
Date: 04/28/20 Time: 22:19
Sample (adjusted): 2002Q2 2017Q4
Included observations: 63 after adjustments

| Variable | Coefficient | Std. Error | t-Statistic | Prob. |
|---|---|---|---|---|
| LOG(OGDIV(-1)) | 0.260254 | 0.095813 | 2.716263 | 0.0087 |
| LOG(ECSTRUCTURECONV) | -2.183034 | 0.517614 | -4.217490 | 0.0001 |
| LOG(OPENNESS) | -2.217883 | 0.931531 | -2.380901 | 0.0206 |
| LOG(WAGESTRUCTURECONV) | -3.049311 | 0.946083 | -3.223091 | 0.0021 |
| LOG(REALGDPCAPCONV) | 3.071194 | 0.905555 | 3.391507 | 0.0013 |
| C | 21.73555 | 4.341481 | 5.006483 | 0.0000 |

| | | | |
|---|---|---|---|
| R-squared | 0.621560 | Mean dependent var | -0.047655 |
| Adjusted R-squared | 0.588364 | S.D. dependent var | 0.933847 |
| S.E. of regression | 0.599146 | Akaike info criterion | 1.903769 |
| Sum squared resid | 20.46161 | Schwarz criterion | 2.107877 |
| Log likelihood | -53.96873 | Hannan-Quinn criter. | 1.984046 |
| F-statistic | 18.72369 | Durbin-Watson stat | 2.161318 |
| Prob(F-statistic) | 0.000000 | | |

**Source:** Own calculations using Eviews 9.0.

As can be seen in Figure 3, five estimators are significant at a threshold of 1%, while the coefficient of economic openness is significant at 5%. The selected independent variables explains the evolution of the business cycles divergence between Romania and Euro Area in a proportion of 62.15% according to the estimated value of R-squared. In addition, the probability of Fisher test confirms the statistical validity of the model, but the confirmation of the maximum verisimilitude of the estimators also imply the need to check the residuals, stability of the model and the existence of multicollinearity. Table 2 provide the main results of the tests used to investigate the residuals features. Thus, the probability of Jarque-Bera confirmed the normal distribution of the residuals, this being an important assumption of the Gauss-Markov theorem. Moreover, the absence of the autocorrelation between residuals was validated by the Breusch-Godfrey test, while Breusch-Pagan-Godfrey confirmed the hypothesis of homoskedasticity.

**Table 2**. *Statistical data series used*

| Hypothesis checked – test peformed | Probability |
|---|---|
| Normal distribution of the residuals<br>- Jarque-Bera test | 0.669 |
| Autocorrelation of the residuals<br>- Breusch-Godfrey test (2 lags included) | 0.604 |
| Homoskedasticity<br>- Breusch-Pagan-Godfrey | 0.181 |

**Source:** Own calculations using Eviews 9.0.

The stability of the model was also confirmed according to the Figure 4, since CUSUM test provides a result significant at 5%. Lastly, I performed the Variance Inflation Factors test (Figure 5) in order to check the existence of multicollinearity. The test provided adequate results given that centered variance inflation factors are lower than 4, this being the accepted threshold for the existence of this issue in the economic community. However, in certain conditions multicollinearity can be accepted if variance inflation factors are higher than 4 and lower than 10, but the values exceeding 10 confirmed the hypothesis of a severe multicollinearity. Taking into consideration all hypothesis analysed, I confirmed that the model provides accurate estimators and there is not any open issues that can affect the reliability of the estimation.

**Figure 4.** *Stability of the model – CUSUM test*

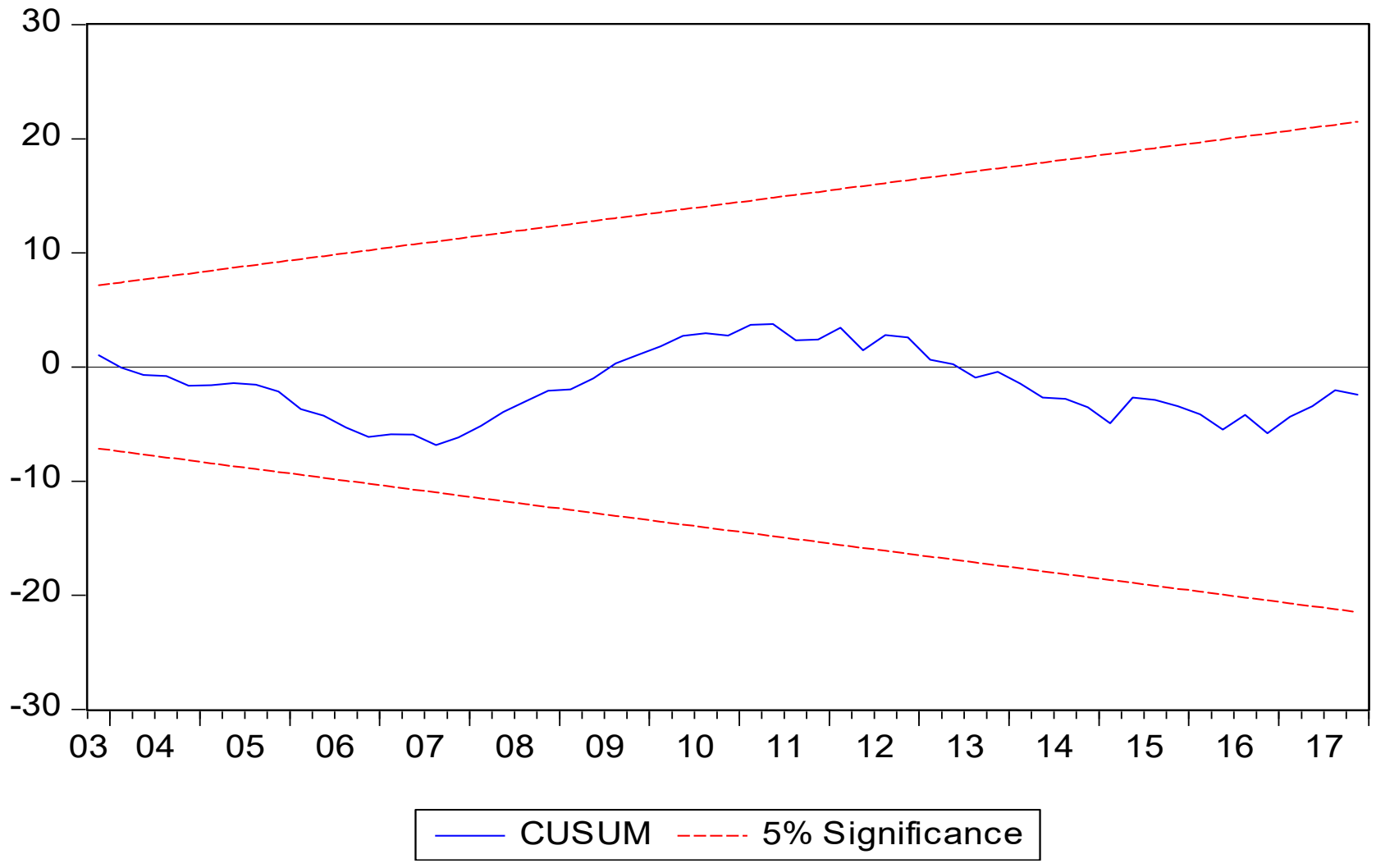

**Source:** Own calculations using Eviews 9.0.

**Figure 5.** *Multicollinearity test – Variance Inflation Factors*

Variance Inflation Factors
Date: 04/28/20 Time: 22:30
Sample: 2002Q1 2017Q4
Included observations: 63

| Variable | Coefficient Variance | Uncentered VIF | Centered VIF |
|---|---|---|---|
| LOG(OGDIV(-1)) | 0.009180 | 1.647412 | 1.632927 |
| LOG(ECSTRUCTUR... | 0.267925 | 756.5723 | 1.379748 |
| LOG(OPENNESS) | 0.867750 | 2869.413 | 1.451064 |
| LOG(WAGESTRUCT... | 0.895072 | 2749.195 | 3.488934 |
| LOG(REALGDPCAPC... | 0.820029 | 1341.467 | 3.727275 |
| C | 18.84846 | 3307.892 | NA |

**Source:** Own calculations using Eviews 9.0.

## 5. Conclusions

This paper analyses the evolution of the business cycles divergence between Romania and Euro Area, respectively the effects of its driving forces. According to the results, the correlation of the business cycles entered on a downward trend in the period 2010-2017 (30.10%), compared to the first subperiod analysed (2002-2009, 67.10%) which indicates a fall in the output-gaps convergence between Romania and Euro Area. Generally, this was the result of the different magnitude of the crisis and of the heterogeneity in government' actions. The business cycles synchronisation still stands at a modest level (2002-2017, 59.09%) and confirms that Romania is not prepared to adopt Euro currency in the near future. Of course, with some significant efforts, this process could be achieved in the upcoming years, but with higher costs than benefits.

The results also indicates a negative relationship between three exogenous factors (economic and wage structure convergence, as well as economic openness) and the dependent variable. Moreover, the output-gaps divergence between the parts reviewed is positively driven by the autoregressive term and the share of the Romanian real GDP per capita in the one registered by the Euro Area. However, we should not look at this estimate as a need to make a trade-off between the catching-up process and business cycles synchronisation. The government should monitor at granular level the driving forces of these variables and try to identify a relevant list of factors that can improve both types of convergence (output-gaps convergence and GDP per capita convergence), such as competitive economic sectors, quality of institutions, foreign direct investments and so on.

**Annex 1.** *Real GDP, Potential GDP and Output-gap calculation (Romania and Euro Area)*

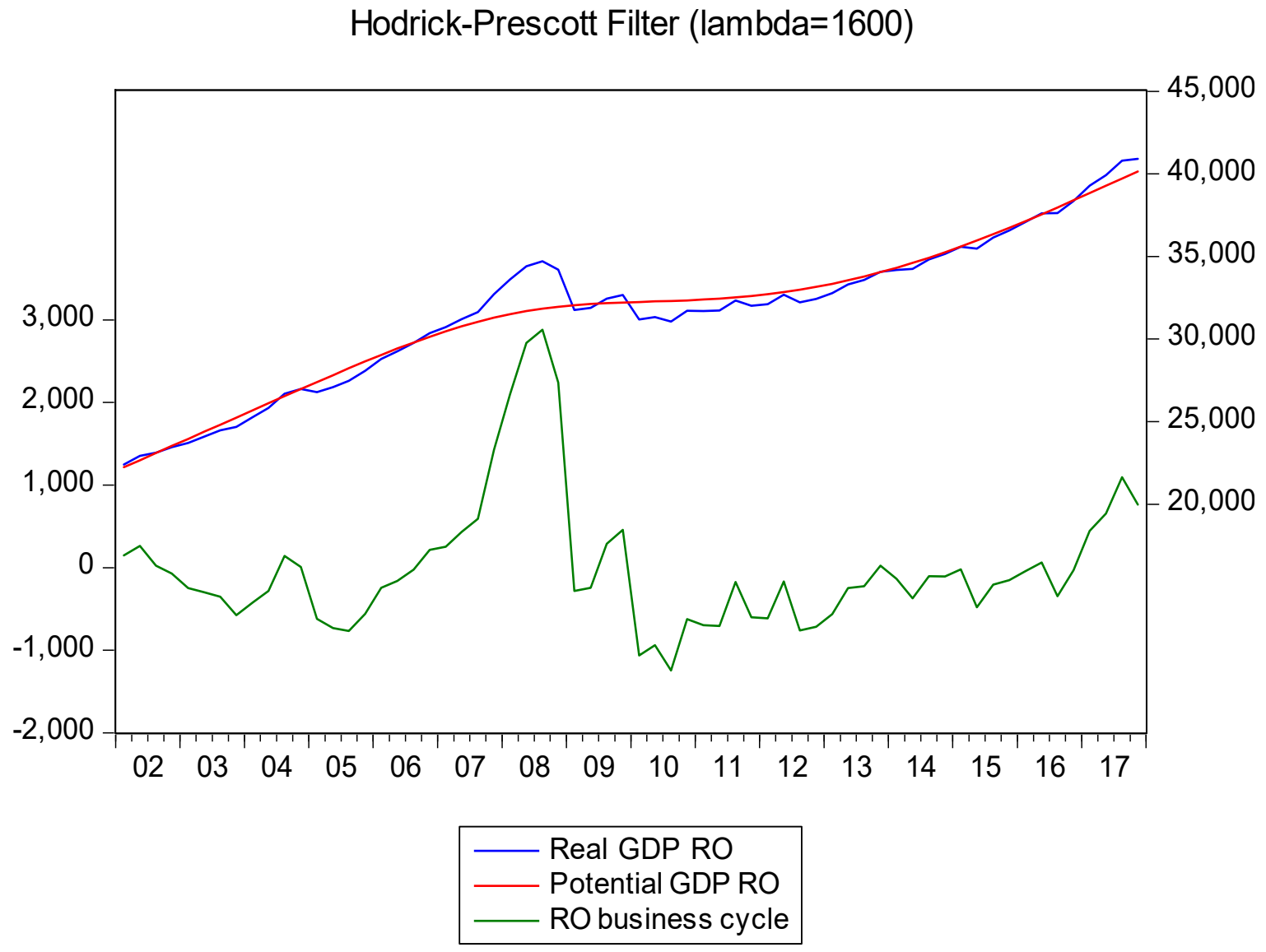

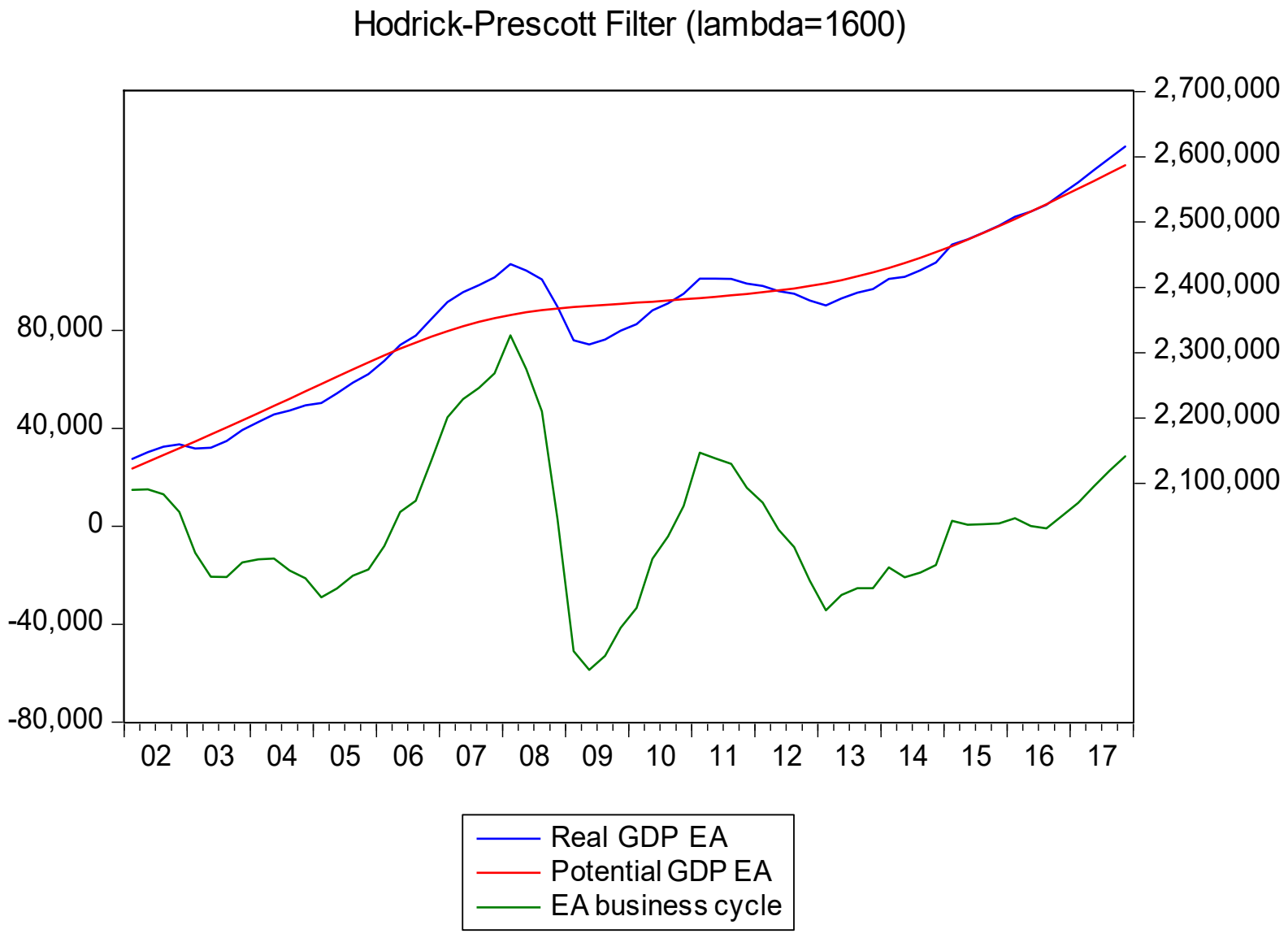

**Source:** Own calculations using Eviews 9.0.